# CLUSTERING PROPERTIES OF THE LYα CLOUDS

S. Cristiani [1], S. D'Odorico[2], V. D'Odorico[1], A. Fontana [3], E. Giallongo [3] and S. Savaglio[4]
[1] *Dipartimento di Astronomia dell'Università, Padova, Italy.*
[2] *European Southern Observatory, Garching, Germany.*
[3] *Osservatorio Astronomico di Roma, Monteporzio, Italy.*
[4] *Dipartimento di Fisica dell'Università, Cosenza, Italy*

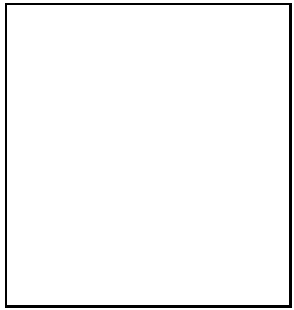

**Abstract**

The spectra of the QSOs PKS2126-158 ($z_{em} = 3.27$) and Q0055-269 ($z_{em} = 3.66$) have been observed at high resolution ($2.2 - 2.7 \times 10^4$). Significant clustering, with $\xi \simeq 1$ at $\Delta v = 100$ km s$^{-1}$, is detected in both cases for lines with $\log N_{HI} \geq 13.8$. A similar result is obtained analysing the data available in the literature for Q0014+81. An anticlustering signal is also detected on scales $\Delta v \sim 600$ km s$^{-1}$. Two voids of size $\sim 2000$ km s$^{-1}$ are found in the spectrum of Q0055-269 with a probability of $2 \times 10^{-4}$.

## 1 Clustering properties of the Ly$_\alpha$ lines

No clustering in the velocity space has been detected so far for Ly$_\alpha$ lines on scales $300 < \Delta v < 30000$ km s$^{-1}$ ([11], [12], [15]). Their spatial distribution is different from that of the metal-line systems selected by means of the CIV doublet, that are known to cluster on scales at least up to 600 km s$^{-1}$ ([13]). Preliminary results at high resolution seem to indicate weak clustering on smaller scales ($\Delta v = 50 - 300$ km s$^{-1}$, [14], [1]).

Although clustering is expected to decrease with redshift, from an empirical point of view it is attractive to push the observations towards high redshifts: the larger the density of lines, the better the chances of detecting a clustering signal. For this reason we have chosen two relatively high-z QSOs, PKS216-158 ($z_{em} = 3.27$) and Q0055-269 ($z_{em} = 3.66$), to be observed at ESO La Silla with the NTT telescope and the EMMI instrument in the echelle mode ([2], [5] and unpublished observations). The weighted mean of the flux-calibrated spectra have a resolution $R \sim 27000$ and $R \sim 22000$ for PKS2126 and Q0055, respectively. The signal-to-noise ratio per resolution element at the continuum level ranges from $S/N \sim 10$ to 40 in the region between Ly$_\alpha$ and Ly$_\beta$. Column densities and Doppler widths of the Ly$_\alpha$ lines and metal-line systems have been derived by a fitting procedure ([6]). The statistical distribution of the Doppler parameter for the Ly$_\alpha$ lines is peaked at $b \simeq 23$ km s$^{-1}$. The column density distribution is described by a power-law with a break or cutoff at $\log N_{HI} \simeq 14.5$. A featureless distribution is rejected with a probability of $> 99.9\%$.

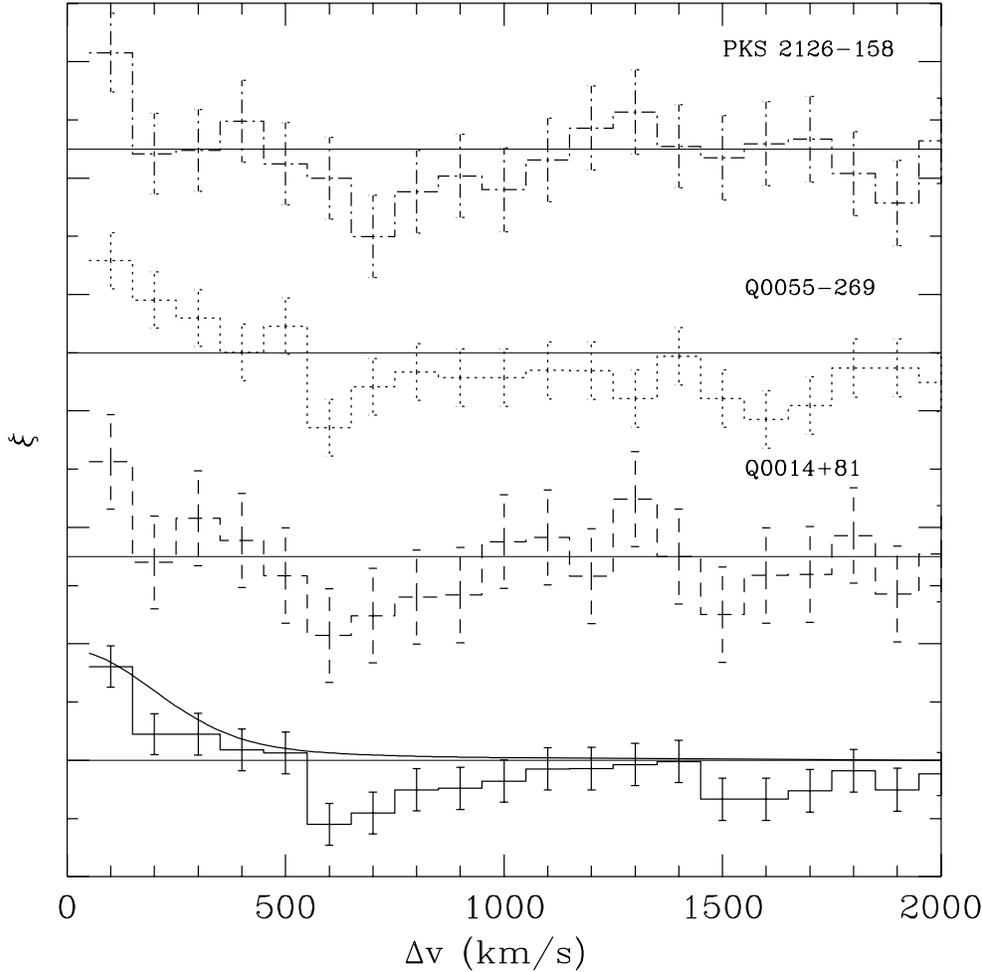

Figure 1: Two-point correlation function of the Lyman$-\alpha$ lines: PKS2126-158 (dot-dashed line), Q0055-269(dotted line), Q0014+81 (dashed line) and weighted mean of the three (continuous line). For each histogram the horizontal continuous line defines the zero level. The distance between two tick marks on the y-axis corresponds to 0.5 units. The continuous curve superimposed to the lowest histogram is the model described in Sec. 2 computed from eq. (2) with $\gamma = 1.77$, $\sigma = 150$ km s$^{-1}$, $r_{cl} = 110$ kpc and $r_o = 280$ kpc at $z = 3$.

We have computed the two-point correlation function in the velocity space for PKS2126-158, Q0055-269 and Q0014+81 ([10]) as:

$$\xi(v, \Delta v) = \frac{N_{obs}(v, \Delta v)}{N_{exp}(v, \Delta v)} - 1 \qquad (1)$$

where $N_{obs}$ is the observed number of line pairs with velocity separations between $v$ and $v + \Delta v$ and $N_{exp}$ is the number of pairs expected from a random distribution in redshift. In our line sample $N_{exp}$ has been obtained averaging 1000 numerical simulations generated according to the cosmological distribution $\propto (1+z)^\gamma$, derived from the maximum likelihood analysis of the real line sample in the appropriate interval of column density and redshift. In this way it is possible to correct for incomplete wavelength coverage due to gaps in the spectra or occultation of weak lines by strong complexes. The region within 8 Mpc from the quasar has been excluded because of the proximity effect. No velocity splittings $\Delta v < 25$ km s$^{-1}$ were included because of the intrinsic line blending due to the typical widths of the Ly$_\alpha$ lines. The 1$\sigma$ deviation from a random distribution has been estimated as $N_{exp}^{-1/2}$, a good approximation in the case of weak clustering $\xi \leq 1$. The resulting correlation function of the Ly$_\alpha$ lines shows a weak but significant signal for all three QSOs on scales $\Delta v \sim 100$ km/s. Exploring the variations of the correlation as a function of the column density, an increase appears as the column density threshold is raised. The maximum significant signal is obtained for lines with $\log N_{HI} \gtrsim 13.8$, for which the measured correlation functions are $\xi_{2126} = 0.83 \pm 0.34$ ($\log N_{HI} \geq 13.85$) $\xi_{0055} = 0.79 \pm 0.24$ ($\log N_{HI} \geq 13.85$), $\xi_{0014} = 0.81 \pm 0.41$ ($\log N_{HI} \geq 14.1$), for $\Delta v = 100$ km s$^{-1}$ (Fig. 1). When computing $\xi$ for lines in the range $\log N_{HI} < 13.6$ (including in the simulated spectra gaps due to lines with $\log N_{HI} > 13.8$), no significant clustering is observed.

The correlation found for the Ly$_\alpha$ cloud positions is less pronounced than for metal-line absorption systems or galaxies, but consistent with a scenario of gravitationally induced correlations, as expected in models where gravitation is an important confining agent ([8]). The correlation found would imply, in the standard CDM framework, that the Ly$_\alpha$ clouds responsible for the weakest absorption are abundant in underdense regions.

Significant anticlustering is detected at scales $\Delta v \sim 600$ km s$^{-1}$ (Fig.1). Possible origins for this phenomenon are inhomogeneities in the UV ionizing background and variations in the baryon to dark matter ratio due to reionization processes (see Bouchet, this meeting).

## 2  The dimensions of the Ly$_\alpha$ clouds

At small velocity differences the three-dimensional spatial autocorrelation function $\xi_r$ and the redshift autocorrelation function are related by the convolution ([7]):

$$\xi_v = \int_0^\infty H dr \; \xi(r) \; P(v \mid r) \propto \int_{r_{cl}}^\infty \frac{H dr}{\sigma} \left(\frac{r}{r_o}\right)^{-\gamma} \left[e^{-\frac{(Hr-v)^2}{2\sigma^2}} + e^{-\frac{(Hr+v)^2}{2\sigma^2}}\right] \qquad (2)$$

where a Gaussian distribution of the peculiar motions with respect to the Hubble flow is assumed together with a power-law spatial correlation function of the galaxy-type.

At small velocity splittings, the redshift correlation scale $r_o$ depends mainly on the cloud sizes $r_{cl}$ and on the velocity dispersion. Although these quantities are poorly known, it is instructive to derive constraints on the cloud sizes and velocity dispersions from the observed correlation. Assuming an index $\gamma = 1.77$ for the power-law spatial correlation function, the fit shown in Fig.1 provides upper limits on the cloud sizes as a function of the velocity dispersion and of the appropriate correlation scale. We obtain $r_{cl} \sim 100 - 190$ kpc for $\sigma_v = 50 - 160$ km s$^{-1}$, and $r_o \sim 245 - 420$ kpc, respectively. In the former case a good fit to the overall shape of the observed correlation function requires a very low value $\gamma = 1.1$. For $\sigma_v > 200$ km s$^{-1}$ the redshift correlation function becomes too flat and very steep $\gamma$ values are needed (e.g. $\sigma_v = 300$ km s$^{-1}$ requires $\gamma = 4$).

## 3  Voids in the Ly$_\alpha$ forest

Voids in the Ly$_\alpha$ forest also provide a test for models of the large-scale structure and of the homogeneity of the UV ionizing flux. Searches for megaparsec-sized voids have produced a couple of claims ([3], [4]), but uncertainties in the line statistics strongly influence the probability estimate ([9]). High resolution data, less affected by blending effects, are ideal in this respect.

We have searched for gaps in a sample of Ly$_\alpha$ lines of Q0055-269 limited, to ensure statistical completeness and accuracy in the deblending, to $\log N_{HI} \geq 13.3$ and $b \leq 30$ km s$^{-1}$. Two regions devoid of such lines have been found, centered respectively at $\lambda \sim 5000, 5206$ Å, with sizes $\Delta v \sim 2009, 2046$ km s$^{-1}$ (i.e. $\sim 20$ Mpc). To establish the random probability of observing such gaps, a set of 50000 spectra have been simulated with the observed number ($N = 178$) of redshifts randomly generated in the same interval of the data (excluding the region within 8 Mpc from the quasar) according to the cosmological distribution. The probability of getting a gap larger than the maximum observed one is then $\simeq 0.02$. The 2046 km s$^{-1}$ gap is significant only to 2.2$\sigma$ level, but the joint probability for the presence of the two observed gaps in the same spectrum is $\simeq 2 \times 10^{-4}$. Even if underdense regions are statistically significant in our spectrum, the filling factor is less than 10%.